# Computing the K-terminal Reliability of Circle Graphs

Min-Sheng Lin[*], Chien-Min Chen

Department of Electrical Engineering

National Taipei University of Technology

Taipei, Taiwan

**Abstract**

Let $G$ denote a graph and let $K$ be a subset of vertices that are a set of target vertices of $G$. The K-terminal reliability of $G$ is defined as the probability that all target vertices in $K$ are connected, considering the possible failures of non-target vertices of $G$. The problem of computing $K$-terminal reliability is known to be #P-complete for polygon-circle graphs, and can be solved in polynomial-time for t-polygon graphs, which are a subclass of polygon-circle graphs. The class of circle graphs is a subclass of polygon-circle graphs and a superclass of t-polygon graphs. Therefore, the problem of computing $K$-terminal reliability for circle graphs is of particular interest. This paper proves that the problem remains #P-complete even for circle graphs. Additionally, this paper proposes a linear-time algorithm for solving the problem for proper circular-arc graphs, which are a subclass of circle graphs and a superclass of proper interval graphs.

## 1. Introduction

Consider a probabilistic graph $G$ in which the edges are perfectly reliable, but vertices may fail with known probabilities. A subset $K$ of vertices of $G$ is chosen as the set of *target vertices* of $G$. The *K-terminal reliability* (KTR) of $G$ is defined as the probability that all target vertices in $K$ are connected.

The KTR problem has been proved to be #P-complete for general graphs and to remain so even for chordal graphs [1]. Valiant [2] defined the class of #P problems as those that involve counting access computations for problems in NP, while the class of #P-complete problems includes the hardest

---

problems in #P. As is well known, all algorithms for solving these problems have exponential time complexity, so the development of efficient algorithms for solving this class of problems is almost impossible. However, considering only a restricted subclass of #P-complete problems can reduce this complexity. Many studies have investigated KTR problems for restricted subclasses of intersection graphs.

A graph *G* is an *intersection graph* of a set of objects if an isomorphism exists between the vertices of *G* and the set of objects such that two vertices are adjacent in *G* if and only if their corresponding objects have a nonempty intersection. Well known special classes of intersection graphs include the following. *Interval graphs* [3] are the intersection graphs of a set of intervals on a line; *proper interval graphs* [3] are interval graphs in which no interval properly contains another; *circular-arc graphs* [3] are the intersection graphs of a set of arcs on a circle; *proper circular-arc graphs* [3] are circular-arc graphs in which no arc properly contains another; *permutation graphs* [3] are the intersection graphs of a set of line segments between two parallel lines; *trapezoid graphs* [4] are the intersection graphs of a set of trapezoids between two parallel lines; *chordal graphs* [3] are the intersection graphs of a set of subtrees of a tree; *circle graphs* [5] are the intersection graphs of a set of chords on a circle; *polygon-circle graphs* [6] are the intersection graphs of a set of convex polygons with corners on a circle; *circle-trapezoid graphs* [7] are the intersection graphs of a set of circle trapezoids on a circle, and *t-polygon graphs* [8] are the intersection graphs of a set of chords in a convex t-sided polygon. Figure 1 depicted the inclusions between these graph classes. Some of the inclusions are trivial: for example, proper interval $\subseteq$ {interval, proper circular-arc} $\subseteq$ circular-arc, interval $\subseteq$ chordal, permutation $\subseteq$ t-polygon $\subseteq$ circle $\subseteq$ circle-trapezoid $\subseteq$ polygon-circle, {interval, permutation} $\subseteq$ trapezoid, and {circle, circular arc, trapezoid} $\subseteq$ circle-trapezoid. However, some of them are non-trivial: for example, chordal $\subseteq$ polygon-circle [9], and proper circular-arc $\subseteq$ circle graph [3].

For the KTR problem, polynomial-time algorithms have been found for interval graphs [1], permutation graphs [1], trapezoid graphs [10], circular-arc graphs [11], and t-polygon graphs [12]. A linear-time algorithm for solving the KTR problem for proper interval graphs was given in [13]. The KTR problem is #P-complete for chordal graphs [1] and, therefore, for polygon-circle graphs. The complexity of the KTR problem remains unsolved for circle graphs and circle-trapezoid graphs. This paper reveals that the KTR problem remains #P-complete for circle graphs and circle-trapezoid graphs but that a further restriction to proper circular-arc graphs admits a linear-time solution.

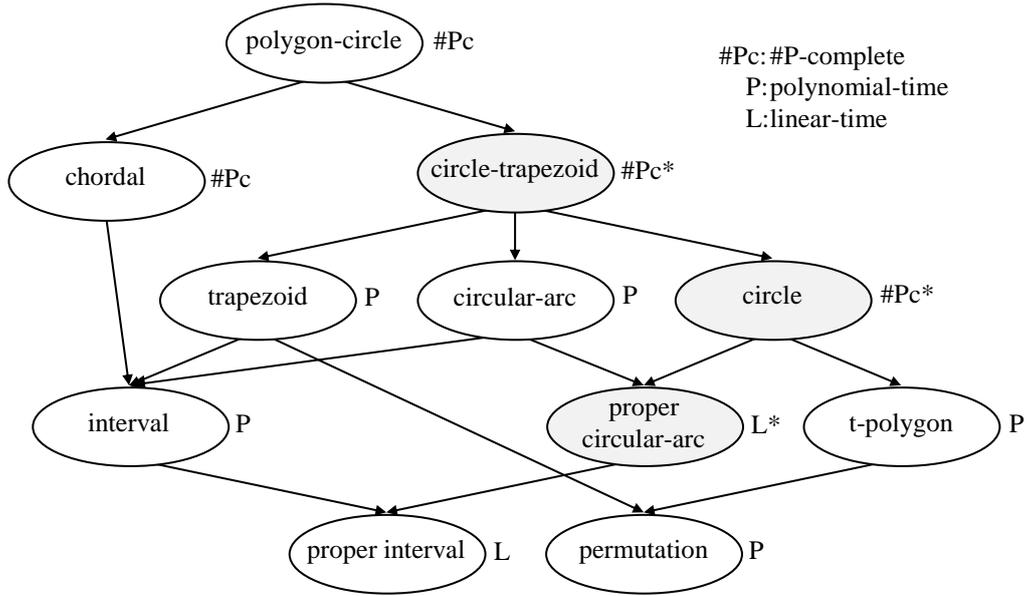

Fig. 1. Inclusion relations between classes of intersection graphs and results of the analysis of the problem of computing KTR. The * symbol indicates a main contribution of this paper.

## 2. Hardness of computing KTR of circle graphs

This section reveals that the KTR problem remains #P-complete for the class of circle graphs. A chord family $\mathcal{C}$ is a set of chords in a circle. A graph $G$ is a circle graph if there exist a chord family $\mathcal{C}$ and a one-to-one mapping between the vertices of $G$ and the chords in $\mathcal{C}$ such that two vertices in $G$ are adjacent if and only if their corresponding chords in $\mathcal{C}$ intersect. Such a chord family $\mathcal{C}$ is called a *circle representation* and $G(\mathcal{C})$ denotes the circle graph that is constructed from $\mathcal{C}$.

For convenience, this paper will consider chords in $\mathcal{C}$ rather than vertices in the corresponding graph $G(\mathcal{C})$. A chord in $\mathcal{C}$ is called a *target chord* if its corresponding vertex in $G(\mathcal{C})$ is a target vertex; otherwise, it is called an *non-target chord*. For simplicity, if no ambiguity arises, let $K$ denote both the set of target chords in $\mathcal{C}$ and the set of target vertices in $G(\mathcal{C})$.

**Theorem 1.** The KTR problem for circle graphs is #*P*-complete.

**Proof.** The KTR problem trivially belongs to #*P*. To prove its #*P*-hardness, the problem of counting edge covers in a bipartite graph, which is known to be #*P*-complete [14], is reduced to the KTR problem for circle graphs.

Given an instance $B$ of bipartite graphs with bipartition $U=\{u_1, u_2, \ldots\}$ and $V=\{v_1, v_2, \ldots\}$, an edge cover of $B$ is a set of edges $D \subseteq E(B)$ such that each vertex in $U \cup V$ is incident to at least one edge of $D$. In the following steps, the corresponding circle representation $\mathcal{C}$ is constructed from $B$. First, corresponding to each vertex $u_i \in U$, chord $x_i$ is placed on the left half circle such that no two chords $x_i$ intersect. Similarly, corresponding to each vertex $v_j \in V$, chord $y_j$ is placed on the right half circle such that no two chords $y_j$ intersect. Next, corresponding to each edge $(u_i, v_j) \in E(B)$, chord $w_{ij}$ is placed on the circle such that chord $w_{ij}$ intersects both chords $x_i$ and $y_j$. Finally, an additional chord $z$ is placed between the left circle and the right circle and it intersects all chords $w_{ij}$ but neither $x_i$ nor $y_j$. Figure 2 shows an example of the above construction.

Let $X = \{ x_i \mid u_i \in U \}$, $Y = \{ y_j \mid v_j \in V \}$, and $W = \{ w_{ij} \mid (v_i, u_j) \in E(B) \}$. Now, $\mathcal{C} = X \cup Y \cup \{z\} \cup W$. Let $K = X \cup Y \cup \{z\}$ be the set of target chords and $W$ be the set of non-target chords in $\mathcal{C}$. A success set $S$ is defined herein as a subset of $W$ such that all target chords in $K$ are connected to each other when all chords in $S$ work and all chords in $W \setminus S$ fail. Let $SS(\mathcal{C})$ denote the collection of all success sets in $\mathcal{C}$. Let $EC(B)$ denote the collection of all edge covers in $B$. The equality $|SS(\mathcal{C})| = |EC(B)|$ is established as follows.

Let $D \in EC(B)$ be an arbitrary edge cover of $B$ and $S' \subseteq W$ be the corresponding subset of non-target chords of $G$, so $S' = \{ w_{ij} \mid (x_i, y_j) \in D \}$. Since $D$ covers all vertices in $U \cup V$, each chord in $X \cup Y$ intersects at least one chord of $S'$. Furthermore, $z$ intersects all chords in $S'$. Consequently, all target chords in $X \cup Y \cup \{z\}$ are connected to each other by the set of non-target chords in $S'$, and thus $S'$ is a success set of $\mathcal{C}$. On the other hand, let $S \in SS(\mathcal{C})$ be an arbitrary success set of $\mathcal{C}$ and $D' \subseteq E(B)$ be the corresponding subset of edges of $B$, so $D' = \{(u_i, v_j) \mid w_{ij} \in S\}$. Because each target chord in $X \cup Y$ intersects at least one chord of $S$, $D'$ is an edge cover of $B$. Therefore, an isomorphism exists between the edge covers of $B$ and the success sets of $\mathcal{C}$. Accordingly, $|SS(\mathcal{C})| = |EC(B)|$. Then, the correlation between the KTR of $G(\mathcal{C})$, $R_K(G(\mathcal{C}))$, and the number of edge cover in $B$, $|EC(B)|$, is established.

According to the definition of success sets, $R_K(G(\mathcal{C}))$ is given by

$$R_K(G(\mathcal{C})) = \sum_{S \in SS(\mathcal{C})} \left( \prod_{c \in S}(1-q_c) \times \prod_{c \in W \setminus S} q_c \right), \text{ where } q_c \text{ is the failure probability of chord } c.$$

Let all $q_c$ for $c \in W$ in the above equation have the same value $\frac{1}{2}$; now,

$$R_K(G(\mathcal{C})) = \sum_{S \in SS(\mathcal{C})} \left( (\tfrac{1}{2})^{|S|} \times (\tfrac{1}{2})^{|W \setminus S|} \right) = \sum_{S \in SS(\mathcal{C})} \left( (\tfrac{1}{2})^{|W|} \right) = |SS(\mathcal{C})| \times (\tfrac{1}{2})^{|E(B)|}.$$

Since $|SS(\mathcal{C})| = |EC(B)|$, the number of edge cover of the bipartite graph $B$ is expressed as

$$|EC(B)| = R_K(G(\mathcal{C})) \times 2^{|E(B)|}.$$

Therefore, an efficient algorithm for determining the KTR of the circle graph $G(\mathcal{C})$ yields an efficient algorithm for counting edge covers in the bipartite graph $B$. However, as is well known, the latter counting problem is #P-complete, so the KTR problem must also be #P-complete. □

Since the class of circle graphs is a subclass of circle trapezoids and polygon-circle graphs, the following corollaries immediately follow.

**Corollary 1.** The KTR problem for circle-trapezoid graphs is #P-complete.

**Corollary 2.** [1] The KTR problem for polygon-circle graphs is #P-complete.

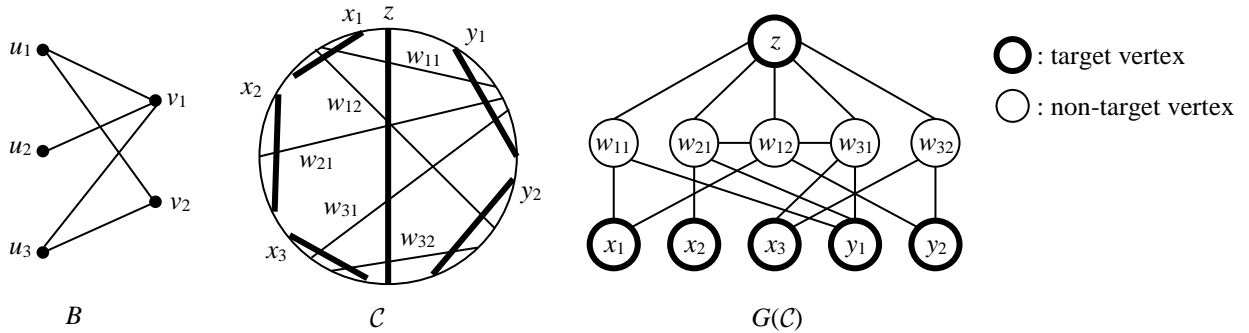

Fig. 2. A bipartite graph $B$, a circle representation $\mathcal{C}$ constructed from $B$, and the circle graph $G(\mathcal{C})$.

## 3. Linear-time algorithm for computing KTR of proper circular-arc graphs

This section presents a linear time algorithm for computing the KTR of proper circular-arc graphs, which are a subclass of circle graphs and a superclass of proper interval graphs. First, some necessary terminology associate with proper circular-arc graphs is introduced; it is similar to that associated with circle graphs in Section 2. A circular-arc family $\mathcal{A}$ is a set of arcs on a circle. A graph $G$ is a circular-arc graph if a circular-arc family $\mathcal{A}$ and a one-to-one mapping between the vertices of $G$ and the arcs in $\mathcal{A}$ exist such that two vertices in $G$ are adjacent if and only if their corresponding arcs in $\mathcal{A}$ intersect. Such a circular-arc family $\mathcal{A}$ is called a *circular-arc representation* for $G$ and $G(\mathcal{A})$ denotes the circular-arc graph constructed from $\mathcal{A}$. If $\mathcal{A}$ can be chosen so that no arc contains another, then $G(\mathcal{A})$ is called a proper circular-arc graph and $\mathcal{A}$ is called a proper circular-arc representation.

As mentioned previously, a polynomial-time algorithm exists for computing the KTR of circular-arc graphs [11]. Since every proper circular-arc graph is a circular-arc graph, the same algorithm can be used to compute the KTR of proper circular-arc graphs. This section presents a more efficient algorithm for computing the KTR of proper circular-arc graphs. Consider a proper circular-arc representation $\mathcal{A}$ with $n$ arcs. For convenience, arcs in $\mathcal{A}$ rather than vertices in $G(\mathcal{A})$ will be considered. An arc in $\mathcal{A}$ is called a *target arc* if its corresponding vertex in $G(\mathcal{A})$ is a target vertex; otherwise, it is called a *non-target arc*.

Starting with an arbitrary arc of $\mathcal{A}$ and traversing the circle clockwise, label all arcs of $\mathcal{A}$ from 0 to $n-1$ in the order in which they are encountered. Let $K=\{x_0 < x_1 < ... < x_{k-1}\}$ be the set of the given $k$ target arcs of $\mathcal{A}$, and let $A_i$ be the set of arcs that are encountered in a clockwise traversal on the circle from arc $x_i$ to arc $x_{i+1}$. Here arithmetic operations on the index $i$ of $x_i$ are taken modulo $k$. That is, $A_i=\{\text{arc } r|\ x_i \leq r \leq x_{i+1}\}$ for $0 \leq i \leq k-2$ and $A_{k-1}=\{\text{arc } r|\ r \geq x_{k-1} \text{ or } r \leq x_0\ \}$. Let $H_i$ be the subgraph of $G(\mathcal{A})$ that is induced by the vertices corresponding to the arcs of $A_i$. The following proposition follows from the definitions of proper interval graphs.

**Proposition 1.** For $0 \leq i \leq k-1$, $H_i$ is a proper interval graph.

Let $R(H_i)$ be the probability that target arcs $x_i$ and $x_{i+1}$ are connected in $H_i$. According to Proposition 1, $H_i$ is a proper interval graph and thus $R(H_i)$ can be computed using the algorithm in our previous work [13], which proposed a linear-time algorithm for computing the KTR of proper interval graphs. Some of the notation used here differs slightly from that used in that work [13], and is described below.

Consider the proper interval graph $H_i$ with the corresponding set of arcs (intervals) $A_i$ on the circle. For each arc $r$ in $A_i$, define $\alpha(r)$ ($\beta(r)$) as the arc $r'$ in $A_i$ such that $r$ and $r'$ intersect and $r'$ is the first (last) arc that is encountered in a clockwise traversal on the circle from $x_i$ to $x_{i+1}$. Also, define $N^+(r)=\{r+1, r+2, \ldots, \beta(r)\}$. Here arithmetic operations on arc $r$ are taken modulo $n$. Now define the event $F(r)$, for $r \in A_i \setminus \{x_{i+1}\}$, as

$$F(r) \equiv \{ \text{ there exists no operating path from arc } x_i \text{ to any arc of } N^+(r) \}.$$

The following lemma provides a recursive method for computing $\Pr[F(r)]$.

**Lemma 1.** [13] For $r \in A_i \setminus \{x_{i+1}\}$,

$$\Pr[F(r)] = \Pr[F(r-1)] + (1 - \Pr[F(\alpha(r)-1)]) \times (1 - q_r) \times \prod_{r'=r+1}^{\beta(r)} q_{r'}, \quad (1)$$

where the boundary condition $Pr[F(x_i-1)]=0$ and $q_r$ is the failure probability of arc $r$.

Notably, since $N^+(x_{i+1}-1)$ is the set of arcs that are encountered in a clockwise traversal from $x_{i+1}$ to $\beta(x_{i+1}-1) \equiv x_{i+1}$, it contains a single arc $x_{i+1}$. Therefore, the event $F(x_{i+1}-1)$ is equivalent to the event $\{$ there exists no operating path from arc $x_i$ to arc $x_{i+1} \}$, and it implies that $Pr[F(x_{i+1}-1)] = 1 - R(H_i)$.

Let $\mathcal{F}$ denote the collection of all events $F(x_{i+1}-1)$, for $0 \leq i \leq k-1$. Some pair of target arcs in $K$ is easily verified to be disconnected if and only if at least two events in $\mathcal{F}$ occur. Let $f(j, h)$, $1 \leq h \leq j \leq k$, be the probability that at least $h$ of the first $j$ events in $\mathcal{F}$ occur. Therefore, the KTR of the proper circular-arc graph is obtained as $1-f(k, 2)$. For simplicity, let $Q_j = \Pr[F(x_j-1)]$ denote the probability that event $F(x_j-1)$ occurs, for $1 \leq j \leq k$. Since the intersection of $A_i$ and $A_{i'}$, for $0 \leq i \neq i' \leq k-1$, contains only perfect target arcs of $K$, any two events in $\mathcal{F}$ are mutually independent. Accordingly, the following recursive relation for computing $f(j, h)$ is easily verified.

**Lemma 2.** For $1 \leq j \leq k$ and $h=1$ or 2,

$$f(j, h) = Q_j \times f(j-1, h-1) + (1-Q_j) \times f(j-1, h), \qquad (2)$$

under the boundary conditions $f(j, 0)=1$, for $0 \leq j \leq k$, and $f(0, 1)= f(0, 2)=0$.

Based on the above formulations, the formal algorithm for computing the KTR of a proper circular-arc graph is presented as follows.

**Algorithm 1**. *Compute the KTR of a proper circular-arc graph*

Input: proper circular-arc representation $\mathcal{A}$ of $n$ arcs labeled clockwise from 0 to $n-1$,

    closed neighborhood structure, $N[r]$, for each arc $r$ of $\mathcal{A}$, and

    set of target arcs $K=\{x_0 < x_1 ...< x_{k-1}\}$ with $k \geq 2$.

Output: KTR of $G(\mathcal{A})$

1   **for** $i \leftarrow 0$ **to** $k-1$ **step** 1 **do begin**
2      $s \leftarrow x_i$; $t \leftarrow x_{i+1 (\text{mod } k)}$; $j \leftarrow i+1$; // for simplicity //
3      **if** ( arcs $s$ and $t$ intersect ) **then**
4         $Q_j \leftarrow 0$;
5      **else begin**
         // find the values of $\alpha(r)$ and $\beta(r)$, for all arc $r$ of $\mathcal{A}$ //
6         $r \leftarrow s$;
7         **while** ( $r \neq t$ ) **do begin**
8            $\alpha(r) \leftarrow r$; $\beta(r) \leftarrow r$; // initial values//
9            **for each** $r' \in N[r]$ **do begin**
10               **if** ( arcs $s$, $r'$, $\alpha(r)$, and $t$ occur in clockwise order ) **then** $\alpha(r) \leftarrow r'$;
11               **if** ( arcs $s$, $\beta(r)$, $r'$ and $t$ occur in clockwise order ) **then** $\beta(r) \leftarrow r'$;
12            **end-for-each**
13            $r \leftarrow r+1 (\text{mod } n)$;
14        **end-while**
         // compute the values of $\Pr[F(r)]$ in Eq.(1) , for each arc $r$ of $\mathcal{A}$ //
15        $PrF[s-1 (\text{mod } n)] \leftarrow 0$; // boundary condition //
16        $r \leftarrow s$;

| 17 | **while** ( $r \neq t$ ) **do begin** |
|---|---|
| 18 | $q^* \leftarrow q_{\beta(r)}$; // initial value // |
| 19 | $r' \leftarrow r+1 \pmod{n}$; |
| 20 | **while** ( $r' \neq \beta(r)$ ) **do begin** |
| 21 | $q^* \leftarrow q^* \times q_{r'}$ ; |
| 22 | $r' \leftarrow r'+1 \pmod{n}$; |
| 23 | **end-while** |
| 24 | $PrF[r] \leftarrow PrF[r-1 \pmod{n}] + (1-PrF[\alpha(r)-1 \pmod{n}]) \times (1-q_r) \times q^*$;   //Eq. (1)// |
| 25 | $r \leftarrow r+1 \pmod{n}$; |
| 26 | **end-while** |
| 27 | $Q_j \leftarrow PrF[t-1 \pmod{n}]$; |
| 28 | **end-if-else** |
| 29 | **end-for** |

// compute the values of $f(j, h)$, for $1 \leq j \leq k$ and $h=1$ or $2$ //

| 30 | **for** $j \leftarrow 0$ **to** $k$ **step** 1 **do** $f(j, 0) \leftarrow 1$;   // boundary conditions // |
|---|---|
| 31 | $f(0, 1) \leftarrow 0$;  $f(0, 2) \leftarrow 0$; // boundary conditions // |
| 32 | **for** $j \leftarrow 1$ **to** $k$ **step** 1 **do begin** |
| 33 |     $f(j, 1) \leftarrow Q_j \times f(j-1, 0) + (1-Q_j) \times f(j-1, 1)$;   // Eq. (2), for $h=1$ // |
| 34 |     $f(j, 2) \leftarrow Q_j \times f(j-1, 1) + (1-Q_j) \times f(j-1, 2)$;   // Eq. (2), for $h=2$ // |
| 35 | **end-for** |
| 36 | **return**( $1-f(k, 2)$ );   // return the KTR of $G(\mathcal{A})$ // |

**end-algorithm**

**Theorem 2.** Given a proper circular-arc graph $G$ with $n$ vertices and $m$ edges, the KTR problem on $G$ can be solved in $O(n+m)$ time.

**Proof.** The correctness of Algorithm 1 follows from above discussion. Notably, if the corresponding proper circular-arc representation $\mathcal{A}$ of the proper circular-arc graph $G$ is not given, then it can be constructed in linear-time using the recognition algorithm [15] for a proper circular-arc graph. First, the values of all $\alpha(r)$ and $\beta(r)$, for $r \in \mathcal{A}$, are obtained by executing lines 10 and 11 at most $O(\sum_{i=0}^{k-1} \sum_{r \in A_i} |N[r]|) = O(n+m)$ times. Next, according to the definition of $\beta(r)$, the while-loop of lines 20 to 23 is iterated at most $|N[r]|$ times for each arc $r$. Therefore, the values of all $PrF[r]$, for $r \in \mathcal{A}$, can be obtained in $O(\sum_{i=0}^{k-1} \sum_{r \in A_i} |N[r]|) = O(n+m)$ time. Finally, the values of all $f(j,1)$ and $f(j,2)$, $1 \leq j \leq k$, can be computed in $O(k)$ time by executing the for-loop of lines 32 to 35. Accordingly, implementing Algorithm 1 takes $O(n+m)$ time overall, which is linear in the size of $G$.

## 4. Conclusions

This paper reveals that the KTR problem remains #P-complete even for circle-graphs. This paper also proposes an efficient linear-time algorithm to solve the KTR problem for proper circular-arc graphs, which are a subclass of circle graphs and a superclass of proper interval graphs. Therefore, the classes of intersection graphs with linear-time solvable KTR problems are extended from proper interval graphs to proper circular-arc graphs.

## Acknowledgment

The authors would like to thank the Ministry of Science and Technology of Taiwan for financially supporting this research under Contract No. MOST 105-2221-E-027-077-.